\documentclass[aps,12pt,superscriptaddress,preprint,showpacs]{revtex4-1}

\usepackage[dvips]{graphicx}
\usepackage{amsmath,amssymb,amsthm,bm}

\renewcommand{\phi}{\varphi} 
\newcommand{\R}{\mathbb{R}} 

\theoremstyle{plain}

\theoremstyle{definition}

\allowdisplaybreaks

\begin{document}

\title{Toward solving the cosmological constant problem by embedding}

\author{Roman V. Buniy}
\email{roman.buniy@gmail.com}
\affiliation{School of Earth and Space Exploration, Arizona State University, Tempe, AZ 85287}
\affiliation{Chapman University, Schmid College of Science, Orange, CA 92866}
\altaffiliation{Permanent address}

\author{Thomas W. Kephart} 
\email{tom.kephart@gmail.com}
\affiliation{Department of Physics and Astronomy, Vanderbilt
  University, Nashville, TN 37235}

\date{\today}

\begin{abstract}
  The typical scalar field theory has a cosmological constant
  problem. We propose a generic mechanism by which this problem is
  avoided at tree level by embedding the theory into a larger theory. The metric and
  the scalar field coupling constants in the original theory do not
  need to be fine-tuned, while the extra scalar field parameters and
  the metric associated with the extended theory are fine-tuned
  dynamically. Hence, no fine-tuning of parameters in the full
  Lagrangian is needed for the vacuum energy in the new physical
  system to vanish at tree level. The cosmological constant problem can be solved if the 
  method can be extended to quantum loops.
\end{abstract} 


\maketitle

\section{Introduction}

There is no symmetry to prevent a   term $\Lambda g_{\mu \nu}$  from being added to the Einstein equation, where $\Lambda$
is called the cosmological constant and  $ g_{\mu \nu}$ is the metric tensor. If this is done 
in a cosmological context, the natural scale for the cosmological constant is $\Lambda \sim m_{\textrm{P}}^4$, where $m_{\textrm{P}}$ is the Planck scale, which disagrees with observation by approximately 120 orders of magnitude.
This disagreement is the cosmological constant problem and it has been with
us for some time now (for reviews see for example
Refs.~\cite{Weinberg:1988cp,Carroll:2000fy,Padmanabhan:2002ji,Nobbenhuis:2004wn,Nobbenhuis:2006yf}). Ideally
we would like a cosmology where the cosmological constant is zero to
first approximation, but corrected by some process to the small value
observed today.

There have been many attempts to solve the cosmological constant
problem in classical and quantum field theory. Early examples include the use of extra dimensions, where it
was conjectured in Ref.~\cite{Rubakov:1983bz} that theories with
$\Lambda \sim 0$ can be picked out by quantum corrections. Application
of the anthropic principle~\cite{Weinberg:1987dv,Weinberg:2000qm} and
backreaction arguments~\cite{Tsamis:1992sx} have also been used to
zero $\Lambda$. It was argued that if wormholes exist then $\Lambda$
can vanish~\cite{Coleman:1988tj}. This led to a long, sometimes
controversial, discussion in the
literature~\cite{Klebanov:1988eh,Banks:1988je,Carlini:1992wh}. A
technically similar, but physically different solution was presented
in Ref.~\cite{Ng:1990rw}, where it was argued that $\Lambda \simeq 0$
dominates the euclidean path integral. It was argued in
Ref.~\cite{Moffat:1996yu} that a stochastic model of vacuum energy
fluctuations treated as a non-equilibrium process gives a natural
explanation for the smallness of $\Lambda$.  Higher spin models have
been introduced to solve the cosmological constant
problem~\cite{Dolgov:1996zg}, an approach that was challenged~\cite{Rubakov:1999bw},
but more recently the objection has been circumvented~\cite{Emelyanov:2011wn}. An interesting interpretation of
cosmological constant problem given in Ref.~\cite{Mannheim:1999nc}
allows a large $\Lambda$ that can be made compatible with
observation. Somewhat closer to the spirit of the present paper are
the works on
$k$-essence~\cite{ArmendarizPicon:2000dh,ArmendarizPicon:2000ah,Barger:2000jg,Chiba:2002mw,Chimento:2003zf,Malquarti:2003nn,Scherrer:2004au,Sen:2005ra,Das:2006cm}.
Finally, the relaxation of boundary and hermiticity constraints on
quantum fields has been shown to have implications for the
cosmological constant problem \cite{'tHooft:2006rs}.

Much work has also been done on the cosmological constant problem in
string and M-theory. In Ref.~\cite{Bousso:2000xa} it was shown that
the cosmological constant can be neutralized by multiple 4-fluxes in
M-theory, braneworld solutions have been given in
Refs.~\cite{Tye:2000fw,Kehagias:2004fb,Diakonos:2007au}, and more
recently, a string theory landscape solution to the cosmological
constant problem~\cite{Douglas:2003um} has generated a considerable
amount of interest.  In addition, various self-tuning mechanisms, with
and without extra dimensions, have been considered
\cite{Kehagias:2000dg,Kim:2001ez,Cline:2002fc,Kim:2002fd,Kim:2003qy,Prokopec:2006yh}. Finally,
a variety of applications of quantum gravity and modifications of
general relativity have been used to address the cosmological constant
problem
\cite{Moffat:2002ju,Carter:2005sd,Afshordi:2008xu,Stefancic:2008zz}.

While this brief and incomplete summary does not cover all the ideas
put forward for solving the cosmological constant problem, we hope it
at least gives a flavor for the ingenuity being expended toward
finding a compelling solution.

The work we will present here, on  extensions of renormalizable particle physics models by embedding in larger theories,  makes technical progress that sheds light on the 
nature of the cosmological constant problem. What remains is to find how these extensions arise naturally in a more fundamental theory. 

We should point out that the method we propose here does not contradict Weinberg's no-go theorem~\cite{Weinberg:1988cp}. The correct counterpart to Weinberg's equation (6.3) is our equation $(\partial {\cal L}/\partial\phi')_*=0$, which is satisfied in our method. However, a related equation $(\partial {\cal L}'/\partial\phi')_*=0$, due to different physical interpretations of $\cal L$ and $\cal L'$, does not need to be satisfied. Similar comments apply to Weinberg's equation (6.2). See below for full details. 

\section{Method} \label{section-method}

For a generic scalar field theory, one reasonably expects that an equilibrium
field configuration is a solution of the equations of motion. It is
likely that the energy of such a configuration is at its minimum, in
which case the solution corresponds to a vacuum state of the
theory. The energy-momentum tensor for such a solution is proportional
to the metric since such a relation is the only one possible given the
symmetry of the vacuum state. The coefficient of proportionality is
the cosmological constant. In order for the cosmological constant to
be zero, parameters of a typical standard field Lagrangian have to be
fine-tuned. For a generic Lagrangian, such fine-tuning depends on the
equilibrium configuration, which makes the occurrence of such a
situation physically highly unlikely.

One possible way to ameliorate  this problem is to introduce extra fields
(whose properties will be elucidated below) in addition to 
fields of the type seen in the standard model of particle physics. These
standard fields are chosen to be in a vacuum state. For each
configuration of the standard field, configurations of the extra
fields must be fine-tuned in such a way as to achieve the vanishing
cosmological constant without fine-tuning of the Lagrangian. If this
fine tuning can be made technically natural, e.g., via dynamics, then
we have solved the cosmological problem.  Since the introduction of
the extra fields changes the energy-momentum tensor, the Einstein
equations require that the space-time metric changes as well. The
original theory will be replaced by an embedding theory with extra
fields, and we require that a suitable projection of the embedding
theory gives the original theory. Clearly, for a given physical
theory, there may be an infinite number of embedding theories. In view of
this, it might be instructive to select certain classes of theories
according to the presence of specific properties, and to choose, based
on certain criteria, the minimal embedding theory. We now give a
mathematical formulation of the method outlined above.

Let $M$ and $N$ be manifolds, $G$ the space of smooth metrics on $M$,
and $\Phi$ the space of smooth maps from $M$ to $N$. We choose a
Lagrangian ${\cal L}:G\times\Phi\to\R$. These objects define a theory
$S$. Let $T$ be the energy-momentum tensor and $E=0$ the equation of
motion for $S$.

Suppose $E(g,\phi)=0$ has a solution $(g,\phi) =(g_*,\phi_*)$, where
$g_*\in G$, $\phi_*\in\Phi$. For an arbitrary quantity $Q(g,\phi)$, we
define $Q_* =Q(g_*,\phi_*)$. Of special interest are solutions for
which $(M,g_*)$ is an Einstein manifold, and we consider only such
solutions. It follows that $(T_*)_{ij} =\Lambda_* (g_*)_{ij}$, where
$\Lambda_*=\textrm{const}$, and we say that $(g_*,\phi_*)$ is a vacuum
solution.

The quantity $\Lambda_*$ plays the role of the cosmological
constant. For a generic ${\cal L}(g,\phi)$, the requirement
$\Lambda_*=0$ leads to the dependence of ${\cal L}$ on
$(g_*,\phi_*)$. In such cases, the theory $S$ has the cosmological
constant problem. The same condition $\Lambda_*=0$ also implies that
$(M,g_*)$ is a Ricci-flat manifold.

Consider the case $N=N'\times N''$, where $N'$ and $N''$ are two
manifolds. Let $\Phi'$ and $\Phi''$ be the spaces of smooth maps from
$M$ to $N'$ and $M$ to $N''$. For an arbitrary quantity $Q$ defined on
$\Phi=\Phi'\times\Phi''$, let $Q'=Q\vert_{\Phi'\times\{0\}}$ and
$Q''=Q\vert_{\{0\}\times\Phi''}$ be the restrictions of $Q$ to
$\Phi'\times\{0\}$ and $\{0\}\times\Phi''$, where $\{0\}$ is the space
of zero functions. As a result, we have the restricted Lagrangians
${\cal L}':G\times\Phi'\to\R$ and ${\cal L}'':G\times\Phi''\to\R$ and
theories $S'$ and $S''$. Let $T'$ and $T''$ be the energy-momentum
tensors and $E'=0$ and $E''=0$ the equations of motion for $S'$ and
$S''$. We say that $S'$ and $S''$ are the sub-theories of $S$ and that
$S$ is the super-theory of $S'$ and $S''$.

Suppose $(g_*,\phi_*)$ is a vacuum solution of $E=0$. We seek vacuum
solutions $(g'_*,\phi'_*)$ and $(g''_*,\phi''_*)$ of $E'=0$ and
$E''=0$ such that $\phi'_*$ and $\phi''_*$ are the restrictions of
$\phi_*$. The quantities $g'_*$ are $g''_*$ are obtained by solving
the equations of motion, not by restricting $g_*$ as the notation may
suggest.

We require $\Lambda_*=0$. Solving the resulting equation $T_*=0$, we
find that ${\cal L}$ depends on $(g_*,\phi_*)$. In general, ${\cal
  L}'$ depends on $(g'_*,\phi'_*)$ and ${\cal L}''$ depends on
$(g''_*,\phi''_*)$. If it is possible to arrange for $\Lambda_*=0$ for
any vacuum solution of $E=0$ in such a way that ${\cal L}'$ does not
depend on $(g'_*,\phi'_*)$ and ${\cal L}''$ does not depend on
$(g''_*,\phi''_*)$, then we say that the cosmological constant
problems for $S'$ and $S''$ are solved by the super-theory $S$.

We will show that the cosmological constant problem for a given
sub-theory $S'$ can always be solved by choosing an appropriate
super-theory. Among all possible super-theories, it might be desirable
to choose a certain super-theory, which is the closest to the given
sub-theory according to some criteria. We call this a minimal
super-theory.

\section{Solution}\label{section-solution}

It is instructive to turn to a specific theory and show how the
general construction described in Section \ref{section-method}
proceeds. We specify $S$ by setting $N'=\R$ and $N''=\R$, and choosing
${\cal L}$ to be an arbitrary $\R$-valued function of $\phi'$,
$\phi''$, $X'$, $X''$, and $Y$, where $\phi'$ and $\phi''$ are real
scalar singlet fields and
\begin{align}
  X' =\nabla_i\phi'\nabla^i\phi', \quad X''
  =\nabla_i\phi''\nabla^i\phi'', \quad Y =\nabla_i\phi'\nabla^i\phi''.
\end{align}
The Euler-Lagrange equations are
\begin{align}
  (\partial {\cal L}/\partial\phi') -\nabla_i \bigl( 2(\partial{\cal
    L}/\partial X')\nabla^i\phi' +(\partial{\cal L}/\partial
  Y)\nabla^i\phi'' \bigr) &=0, \label{EL-equations-S-1}
  \\ (\partial{\cal L}/\partial\phi'') -\nabla_i \bigl( (\partial{\cal
    L}/\partial Y)\nabla^i\phi' +2(\partial{\cal L}/\partial
  X'')\nabla^i\phi'' \bigr) &=0 \label{EL-equations-S-2}
\end{align}
and the energy-momentum tensor is
\begin{align}
  T_{ij} =-{\cal L}g_{ij} +2(\partial {\cal L}/\partial
  X')\nabla_i\phi'\nabla_j\phi' +2(\partial {\cal L}/\partial
  X'')\nabla_i\phi''\nabla_j\phi'' +2(\partial {\cal L}/\partial
  Y)\nabla_i\phi'\nabla_j\phi''. \label{energy-momentum-S}
\end{align}
We seek the solution $(g_*,\phi_*)$ for which $(M,g_*)$ is a
Ricci-flat manifold and $\phi'_*=\textrm{const}$. Equations
\eqref{EL-equations-S-1}, \eqref{EL-equations-S-2},
\eqref{energy-momentum-S} give
\begin{align}
  {\cal L}_* =0, \quad (\partial {\cal L}/\partial\phi')_* =0, \quad
  (\partial {\cal L}/\partial\phi'')_* =0, \quad (\partial {\cal
    L}/\partial X'')_* =0, \quad (\partial {\cal L}/\partial Y)_*
  =0. \label{vacuum-configuration}
\end{align}

We choose ${\cal L}'$ to be an arbitrary $\R$-valued function of
$\phi'$ and $X'$, and ${\cal L}''$ to be an arbitrary $\R$-valued
function of $\phi''$ and $X''$. The Euler-Lagrange equation are
\begin{align}
  (\partial {\cal L}'/\partial\phi') -\nabla_i \bigl( 2(\partial{\cal
    L}'/\partial X')\nabla^i\phi' \bigr) &=0, \\ (\partial {\cal
    L}''/\partial\phi'') -\nabla_i \bigl( 2(\partial{\cal
    L}''/\partial X'')\nabla^i\phi'' \bigr) &=0,
\end{align}
and the energy-momentum tensors are
\begin{align}
  T'_{ij} &=-{\cal L}' g_{ij} +2(\partial {\cal L}'/\partial
  X')\nabla_i\phi'\nabla_j\phi', \\ T''_{ij} &=-{\cal L}'' g_{ij}
  +2(\partial {\cal L}''/\partial X'')\nabla_i\phi''\nabla_j\phi''.
\end{align}
If $(g'_*,\phi'_*)$ and $(g''_*,\phi''_*)$ are vacuum solutions of
$E'=0$ and $E''=0$, then
\begin{align}
  {\cal L}'_* &=(2m)^{-1}(m-2) R'_*, \quad (\partial {\cal
    L}'/\partial\phi')_* =0, \label{conditions-L1} \\ {\cal L}''_*
  &=(2m)^{-1}(m-2) R''_*, \quad (\partial {\cal L}''/\partial\phi'')_*
  =0, \label{conditions-L2}
\end{align}
where $m=\dim{M}$, and $R'_*$ and $R''_*$ are the scalar curvatures of
$(M,g'_*)$ and $(M,g''_*)$. If $S'$ and $S''$ are sub-theories of $S$,
then
\begin{align}
  {\cal L}' ={\cal L}\vert_{\phi''=0, \, X''=0, \, Y=0}, \quad {\cal
    L}'' ={\cal L}\vert_{\phi'=0, \, X'=0, \, Y=0}. \label{limit}
\end{align}

Without loss of generality, we assume that the functions ${\cal L}$,
${\cal L}'$, ${\cal L}''$ can be expanded in power series around the
point $(\phi',\phi'',X',X'',Y)=(0,0,0,0,0)$. Equations \eqref{limit}
imply
\begin{align}
  {\cal L}(\phi',\phi'',X',X'',Y) &={\cal L}'(\phi',X')
  +\sum_{\begin{subarray}{c} p\ge 0, \, q\ge 0, \, r\ge 0 \\ p+q+r \ge
      1 \end{subarray}} F'_{p,q,r}(\phi',X') \phi^{\prime\prime p}
  X^{\prime\prime q} Y^r, \label{expansion-1} \\ {\cal
    L}(\phi',\phi'',X',X'',Y) &={\cal L}''(\phi'',X'')
  +\sum_{\begin{subarray}{c} p\ge 0, \, q\ge 0, \, r\ge 0 \\ p+q+r \ge
      1 \end{subarray}} F''_{p,q,r}(\phi'',X'') \phi^{\prime p}
  X^{\prime q} Y^r, \label{expansion-2}
\end{align}
where $\{F'_{p,q,r}\}$ and $\{F''_{p,q,r}\}$ are arbitrary
functions. Substituting equation \eqref{expansion-1} into equation
\eqref{vacuum-configuration}, we find
\begin{align}
  {\cal L}'_* +\sum_{\begin{subarray}{c} p\ge 0, \, q\ge 0 \\ p+q \ge
      1 \end{subarray}} (F'_{p,q,0})_* \phi_*^{\prime\prime p}
  X_*^{\prime\prime q} &=0, \label{F-equation-1}
  \\ \sum_{\begin{subarray}{c} p\ge 0, \, q\ge 0 \\ p+q \ge
      1 \end{subarray}} (\partial F'_{p,q,0}/\partial\phi')_*
  \phi_*^{\prime\prime p} X_*^{\prime\prime q}
  &=0, \label{F-equation-2} \\ \sum_{\begin{subarray}{c} p\ge 0, \,
      q\ge 0 \\ p+q \ge 1 \end{subarray}} (F'_{p,q,0})_*
  p\phi_*^{\prime\prime p-1} X_*^{\prime\prime q}
  &=0, \label{F-equation-3} \\ \sum_{\begin{subarray}{c} p\ge 0, \,
      q\ge 0 \\ p+q \ge 1 \end{subarray}} (F'_{p,q,0})_*
  \phi_*^{\prime\prime p} q X_*^{\prime\prime q-1}
  &=0, \label{F-equation-4} \\ \sum_{\begin{subarray}{c} p\ge 0, \,
      q\ge 0 \end{subarray}} (F'_{p,q,1})_* \phi_*^{\prime\prime p}
  X_*^{\prime\prime q} &=0. \label{F-equation-5}
\end{align}
Functions $\{F'_{p,q,0}\}$ are constrained by equations
\eqref{F-equation-1}, \eqref{F-equation-2}, \eqref{F-equation-3},
\eqref{F-equation-4}, functions $\{F'_{p,q,1}\}$ are constrained by
equation \eqref{F-equation-5}, and functions $\{F'_{p,q,r}\}$ for
$r\ge 2$ are not constrained by these equations. In general, we assume
that ${\cal L}'_*\not=0$, $\phi''_*\not=0$, $X''_*\not=0$.

Let $k_r$ be the number of nonzero functions among $\{F'_{p,q,r}\}$
for each $r\ge 0$. Equations \eqref{F-equation-1},
\eqref{F-equation-3}, \eqref{F-equation-4} give $(k_0)_*\ge 2$ and
equation \eqref{F-equation-5} gives $(k_1)_*=0$ or $(k_1)_*\ge
2$. Since $k_0\ge (k_0)_*$ and $k_1\ge (k_1)_*$, it follows that
$k_0\ge 2$ and $k_1\ge 0$.

Let $k_0=2$, which implies $(k_0)_*=2$. There are three cases to
consider.

In the first case, $(F'_{p_1,q_1,0},F'_{p_2,q_2,0})\not=(0,0)$, for
some fixed $(p_1,q_1,p_2,q_2)$ such that $p_1\ge 1$, $p_2\ge 1$,
$q_1\ge 1$, $q_2\ge 1$, $(p_1,q_1)\not=(p_2,q_2)$, so that
\begin{align}
  {\cal L} &={\cal L}' +F'_{p_1,q_1,0}\phi^{\prime\prime p_1}
  X^{\prime\prime q_1} +F'_{p_2,q_2,0}\phi^{\prime\prime p_2}
  X^{\prime\prime q_2} +\sum_{\begin{subarray}{c} p\ge 0, \, q\ge 0,
      \, r\ge 1 \end{subarray}} F'_{p,q,r} \phi^{\prime\prime p}
  X^{\prime\prime q} Y^r.
\end{align}
From equations \eqref{F-equation-1}, \eqref{F-equation-3},
\eqref{F-equation-4}, we find $p_1 q_2 =p_2 q_1$ and
\begin{align}
  (F'_{p_1,q_1,0})_* &=s_2(s_1-s_2)^{-1} {\cal L}'_*
  \phi_*^{\prime\prime -p_1} X_*^{\prime\prime -q_1}, \label{F1}
  \\ (F'_{p_2,q_2,0})_* &=s_1(s_2-s_1)^{-1} {\cal L}'_*
  \phi_*^{\prime\prime -p_2} X_*^{\prime\prime -q_2}, \label{F2}
\end{align}
where either $(s_1,s_2)=(p_1,p_2)$ or $(s_1,s_2)=(q_1,q_2)$. Equation
\eqref{F-equation-2} becomes
\begin{align}
 s_1^{-1}(F'_{p_1,q_1,0})_*^{-1}(\partial
 F'_{p_1,q_1,0}/\partial\phi')_*
 -s_2^{-1}(F'_{p_2,q_2,0})_*^{-1}(\partial
 F'_{p_2,q_2,0}/\partial\phi')_* =0. \label{F3}
\end{align}
Since $\phi'_*$ is an arbitrary constant which satisfies only the
condition $(\partial {\cal L}'/\partial\phi')_* =0$, equation
\eqref{F3} implies
\begin{align}
(\partial F'_{p_1,q_1,0}/\partial\phi')_* =0, \quad (\partial
  F'_{p_2,q_2,0}/\partial\phi')_* =0. \label{F4}
\end{align}

In the second case, $(F'_{p_1,0,0},F'_{p_2,0,0})\not=(0,0)$, for some
fixed $(p_1,p_2)$ such that $p_1\ge 1$, $p_2\ge 1$, $p_1\not=
p_2$. The corresponding expressions are obtained from equations
\eqref{F1}, \eqref{F2}, \eqref{F3}, \eqref{F4} by setting
$(q_1,q_2)=(0,0)$ and $(s_1,s_2)=(p_1,p_2)$.

In the third case, $(F'_{0,q_1,0},F'_{0,q_2,0})\not=(0,0)$, for some
fixed $(q_1,q_2)$ such that $q_1\ge 1$, $q_2\ge 1$, $q_1\not=
q_2$. The corresponding expressions are obtained from equations
\eqref{F1}, \eqref{F2}, \eqref{F3}, \eqref{F4} by setting
$(p_1,p_2)=(0,0)$ and $(s_1,s_2)=(q_1,q_2)$.

It is straightforward to proceed with a similar analysis for $k_0\ge
3$.

We require that $\phi''$ is a dynamical field and that $\phi'$ and
$\phi''$ are coupled. If $k_r=0$ for all $r\ge 1$, these conditions
imply $(q_1,q_2)\not=(0,0)$ and
$(F'_{p_1,q_1,0},F'_{p_2,q_2,0})\not=(\text{const},\text{const})$.

\section{Examples}\label{section-examples}

In this section, we restrict our attention to four-dimensional
space-time manifolds, i.e., $m=4$. We define dimensions
\begin{align}
  d(\phi') =1, \quad d(\phi'') =1, \quad d(X') =4, \quad d(X'')=4, \quad
  d(Y) =4
\end{align}
and the corresponding dimension $d(Q)$ of an arbitrary polynomial
$Q(\phi',\phi'',X',X'',Y)$ as the maximal dimension of its
monomials. As a criterion for a minimal super-theory $S$, we choose a
requirement that $d({\cal L})$ takes its least possible value. For
equation \eqref{expansion-1}, we find
\begin{align}
  d({\cal L}) &=\max{\{d({\cal L}'),d'\}}, \\ d'
  &=\max{\{d(F'_{p,q,r}) +p+4(q+r): p\ge 0, \, q\ge 0, \, r\ge 0, \,
    p+q+r \ge 1, \, F'_{p,q,r}\not=0 \}}.
\end{align}
We assume that $d({\cal L}')$ is fixed and thus we need to find the
least possible value for $d'$. 

If $k_0=2$, $k_r=0$, $r\ge 1$, then the least possible value for $d'$
is achieved for
\begin{align}
  (p_1,q_1)=(0,1), \quad (p_2,q_2)=(0,2), \\ 0\le d(F'_{0,1,0}) \le 4,
  \quad d(F'_{0,2,0}) =0, \label{example1-d}\\ {\cal L} ={\cal L}'
  +F'_{0,1,0} X^{\prime\prime} +F'_{0,2,0} X^{\prime\prime
    2}, \label{example1-L} \\ (F'_{0,1,0})_* =-2{\cal L}'_*
  X^{\prime\prime -1}_*, \quad (F'_{0,2,0})_* ={\cal L}'_*
  X^{\prime\prime -2}_*, \label{example1-F*} \\ d({\cal L})
  =\max{\{d({\cal L}'),8\}}.
\end{align}
If $d({\cal L}')>4$, then there does not exist $F'_{0,1,0}$ which
satisfies equations \eqref{example1-d} and \eqref{example1-F*}. If
$0\le d({\cal L}')\le 4$, then
\begin{align}
  F'_{0,1,0} =-2({\cal L}' +C'_1 X') X^{\prime\prime -1}_*, \quad
  F'_{0,2,0} =C'_2, \label{example1-F}
\end{align}
where $C'_1$ and $C'_2$ are arbitrary constants. In Table \ref{table},
we have listed examples which give few smallest values for
$\min{\{d({\cal L} -{\cal L}')\}}$ for the case $k_0=2$, $k_r=0$, $r\ge
1$.

If $k_0=2$, $k_1=1$, $k_r=0$, $r\ge 2$, then the least possible value
for $d'$ is achieved for
\begin{align}
  1\le p_1<p_2\le 4, \quad (q_1,q_2)=(0,0), \\ 0\le d(F'_{p_1,0,0})
  \le 8-p_1, \quad 0\le d(F'_{p_2,0,0}) =8-p_2, \quad d(F'_{0,0,1})
  =4, \label{example2-d} \\ {\cal L} ={\cal L}' +F'_{p_1,0,0}
  \phi^{\prime\prime p_1} +F'_{p_2,0,0} \phi^{\prime\prime p_2}
  +F'_{0,0,1} Y \label{example2-L} \\ (F'_{p_1,0,0})_*
  =p_2(p_1-p_2)^{-1} {\cal L}'_* \phi_*^{\prime\prime -p_1}, \quad
  (F'_{p_2,0,0})_* =p_1(p_2-p_1)^{-1} {\cal L}'_* \phi_*^{\prime\prime
    -p_2}, \label{example2-F*} \\ d({\cal L}) =\max{\{d({\cal
      L}'),8\}}.
\end{align}
We find
\begin{align}
   F'_{p_1,0,0} =p_2(p_1-p_2)^{-1} \bigl( {\cal L}' +C'_1(\phi')X'
   \bigr) \phi_*^{\prime\prime -p_1}, \\ F'_{p_2,0,0}
   =p_1(p_2-p_1)^{-1} \bigl( {\cal L}' +C'_2(\phi')X' \bigr)
   \phi_*^{\prime\prime -p_2}, \label{example2-F}
\end{align}
where $C'_1$ and $C'_2$ are arbitrary polynomials of $\phi'$ such that
\begin{align}
  0\le d(C'_1)\le 4-p_1, \quad 0\le d(C'_2)\le 4-p_2.
\end{align}

\begin{table}
\caption{\label{table} Choices of $(p_1,q_1,p_2,q_2)$ which give few
smallest values for $\min{\{d({\cal L} -{\cal L}')\}}$ for the case
  $k_0=2$, $k_r=0$, $r\ge 1$.  For each choice,
  $(F'_{p_1,q_1,0},F'_{p_2,q_2,0})\not=(\text{const},\text{const})$.}
\begin{ruledtabular}
\begin{tabular}{cccccccc}
  $p_1$ & $q_1$ & $p_2$ & $q_2$ & $\min{\{d({\cal L} -{\cal L}')\}}$ &
  ${\cal L} -{\cal L}'$ & $(F'_{p_1,q_1,0})_*$ & $(F'_{p_2,q_2,0})_*$
  \\ \hline
  $0$ & $1$ & $0$ & $2$ & $8$ & 
  $F'_{0,1,0} X'' 
  +F'_{0,2,0} X^{\prime\prime 2}$ &
  $-2{\cal L}'_* X^{\prime\prime -1}_*$ & 
  ${\cal L}'_* X^{\prime\prime -2}_*$ \\
  $1$ & $1$ & $2$ & $2$ & $10$ & 
  $F'_{1,1,0} \phi^{\prime\prime } X'' 
  +F'_{2,2,0} \phi^{\prime\prime 2} X^{\prime\prime 2}$ &
  $-2{\cal L}'_* \phi^{\prime\prime -1}_*X^{\prime\prime -1}_*$ & 
  ${\cal L}'_* \phi^{\prime\prime -2}_*X^{\prime\prime -2}_*$ \\
  $0$ & $1$ & $0$ & $3$ & $12$ & 
  $F'_{0,1,0} X'' 
  +F'_{0,3,0} X^{\prime\prime 3}$ &
  $-\frac{3}{2}{\cal L}'_* X^{\prime\prime -1}_*$ & 
  $\frac{1}{2}{\cal L}'_* X^{\prime\prime -3}_*$ \\
  $0$ & $2$ & $0$ & $3$ & $12$ & 
  $F'_{0,2,0} X^{\prime\prime 2} 
  +F'_{0,3,0} X^{\prime\prime 3}$ &
  $-3{\cal L}'_* X^{\prime\prime -2}_*$ & 
  $2{\cal L}'_* X^{\prime\prime -3}_*$ \\
  $2$ & $1$ & $4$ & $2$ & $12$ & 
  $F'_{2,1,0} \phi^{\prime\prime 2} X'' 
  +F'_{4,2,0} \phi^{\prime\prime 4} X^{\prime\prime 2}$ &
  $-2{\cal L}'_* \phi^{\prime\prime -2}_*X^{\prime\prime -1}_*$ & 
  ${\cal L}'_* \phi^{\prime\prime -4}_*X^{\prime\prime -2}_*$ \\
\end{tabular}
\end{ruledtabular}
\end{table}

It is straightforward to proceed with a similar analysis for different
values of $\{k_r\}_{r\ge 0}$. The above computations give explicit
construction of the minimal super-theory for a given sub-theory. It is
easy to generalize these computations to more complicated cases such
as, for example, higher dimensional space-times, multiple scalar
fields, or scalar fields in a representation of a gauge group.

\section{Discussion and conclusion}

Here we focus on simple examples of obtaining a zero cosmological
constant. If we have a scalar field theory of the type found as a
component of the standard model of particle physics, where the scalar
field is renormalizable, then we are dealing with the case
\begin{align}
  {\cal L'}(\phi',X') =\tfrac{1}{2}X'-V'(\phi'),\label{phi4}
\end{align}
where $V'(\phi')$ is a polynomial potential of at most dimension $4$.
We call these $\phi'$ standard scalar fields. Note that ${\cal L'}$ is the most 
general renormalizable Lagrangian for a single scalar field.
Unless fine tuned, such
theories always have a cosmological constant problem.  Solutions to
the cosmological constant problem involve fields with nonstandard
properties, which we collectively call exotic scalar fields.  The
class of exotics includes ghost fields with wrong sign kinetic energy
terms, $k$-essence fields with kinetic and potential parts appearing
in the form $ X^{\prime n} V'(\phi')$, phantom fields, auxiliary
fields, and all other types of scalar fields that do not fit the
standard scalar field classification.  The solutions we have found are
models that have a limiting case where only standard fields are
present and where exotic fields must be included to solve the
cosmological constant problem.

For example, ${\cal L'}(\phi',X')$ can be supplemented with an extra
field $\phi''$, the quantity $X''=\nabla_i\phi''\nabla^i\phi''$, and
the new Lagrangian ${\cal L}(\phi',\phi'',X',X'')$ such that there is
no cosmological constant problem for a solution
$(\phi'_*,X'_*)=(\text{const},0)$ with an appropriate choice of
$(\phi''_*,X''_*)$. In the limit $(\phi'',X'')\to (0,0)$, we recover
${\cal L'}(\phi',X')$, and the cosmological constant problem. Note
that the field $\phi'$ itself becomes exotic because of the way
$\phi''$ has to be added to the theory.

Assuming no contributions from the kinetic cross term $Y$, it turns
out that Lagrangians linear in $X''$ are insufficient, but there
exists an infinite class of Lagrangians quadratic in $X''$ which allow
satisfactory solutions. Many of these solutions are in the spirit of a
generalized $k$-essence in the sense that the potential $V'(\phi')$
couples to $X''$. (In $k$-essence, $V'(\phi')$ couples to $X'$.)

Generalizations to models with multiple fields, higher order terms in
$X''$ and $Y$, or more complicated $\phi''$ terms are straightforward.

There is no solution of the cosmological constant problem with
standard fields alone. Any generic standard scalar field Lagrangian is
plagued with a cosmological constant problem and exotic fields are
required to avoid it. We can express this in a concise way since the
results of Sections \ref{section-solution} and \ref{section-examples}
establish that the cosmological constant problem in a standard field Lagrangian can
 only be avoided in a technically natural way by incorporating exotic
 fields of the type introduced above. There are a large variety of exotic field
 properties, including non-polynomial potentials, non quadratic kinetic terms, mixed kinetic-potential terms, etc.

Our method is generic in the sense that for any standard field
Lagrangian there are infinitely many choices for the Lagrangian of the
full system. Since we have a large class of models without a
cosmological constant problem, it is not unrealistic to hope that some
members of the class may arise naturally in a more fundamental
context, like string or M-theory.  Since the way the exotic fields
enter can vary greatly, our results provide a large parameter space of
new models to explore.

Let us consider two simple explicit examples for the new
Lagrangian. In the first example,
\begin{align}
  {\cal L}(\phi',\phi'',X',X'') =\tfrac{1}{2}X'-V'(\phi')(1
  -M^{-4} X'')^2,
\end{align}
and in the second example,
\begin{align}
  {\cal L}(\phi',\phi'',X',X'') =\tfrac{1}{2}X'-\bigl( V'(\phi')^{1/2}
  -M^{-2} X'' \bigr) ^2,
\end{align}
where $M$ is a quantity with the dimension of mass and $V'(\phi')\ge
0$. Note that in neither example do we have symmetry or renormalizability 
to restrict the form of the extended Lagrangian. While these examples do solve the 
cosmological constant problem, for their forms to arise in a natural way we need 
them to be embeddable in an
overarching theory (e.g., string theory) to make that specification. Hence our results
should be considered as technical progress toward a solution of the cosmological 
constant problem until an all-encompassing theory can be found where our examples
can reside.

The solution $\phi'_*=\text{const}$, $X'_*=0$, $X''_*=M^4$ in the
first example and $\phi'_*=\text{const}$, $X'_*=0$, $X''_*=M^2
V'(\phi'_*)^{1/2}$ in the second example solve the Euler-Lagrange
equations for the new Lagrangian. In both examples we see there is no
cosmological constant problem as the value of the energy-momentum
tensor vanishes at the extremum. In the limit $(\phi'',X'')\to (0,0)$,
we recover ${\cal L}'$ of equation \eqref{phi4} from ${\cal L}$. (We
note that in order for the Lagrangian in the second example to agree
with the approach in Section \ref{section-examples}, we need to assume
that the polynomial $V'(\phi')$ is the square of a second order
polynomial of $\phi'$.)

If ${\cal L'}$ is renormalizable and contains only operators of
dimension not exceeding $4$, then the solutions of the cosmological
constant problem we have found in the form of ${\cal L}$ are all
non-renormalizable with operators of at least dimension $8$. If $V'$
has dimension $4$, then ${\cal L}$ has dimension $12$ in the first
example and $8$ in the second example. We assume $M$ is a high scale,
say $M_{\text{GUT}}$ or $M_{\text{Planck}}$, and that the potential in
${\cal L'}$ contains a lower scale $m$, say the electroweak scale in
the form of a mass term in $V'(\phi')= C -m^2\phi^{\prime 2} +\lambda
\phi^{\prime 4}$, where $\lambda \lesssim O(1)$. At temperatures
below, say $1$ TeV, the $\phi'$ potential becomes
\begin{align}
  V'(\phi',T) =C +(-m^2 +\tfrac{1}{2}\lambda T^2)\phi^{\prime 2}
  +\lambda \phi^{\prime 4},
\end{align}
In the first example $X''_*(T)=M^4$ is approximately constant since $M
\gg 1$ TeV and so the $\phi''$ field is ``frozen'' at $T \sim 1$ TeV,
while in the second example,
\begin{align}
  X''_*(T) =M^2 \bigl( C -(4\lambda)^{-1}(-m^2 +\tfrac{1}{2}\lambda
  T^2)^2 \bigr)^{1/2}.
\end{align}
so $\phi''$ is still running with temperature. One concludes that the
solutions to the cosmological problem derived by our methods can have
dramatically different phenomenologies.

As we have pointed out, operators of dimension greater than $4$ are
not surprising from the perspective of string theory, in fact they are
ubiquitous. Hence it would be expected that the high energy completion
of a standard model type Lagrangian involve such operators. As we have
shown, any renormalizable models with a cosmological constant problem
has an infinite class of extensions that solve this problem. Thus it
is quite conceivable that some of these extremal solutions coincide
with members of the vast landscape of string vacua.  To find such a
solution we need not explore the entire string theory landscape
statistically, rather we only need to search for string
compactification with the properties specified above.

Hence we have provided a scenario by which a renormalizable quantum field theory may be extended to solve the 
cosmological constant problem. While it seems unlikely that this solution can
withstand all possible scrutiny, we do believe we have made progress in finding a
deeper understanding of the problem and hope our work will spark further discussion.
Assuming there exists a viable UV completion of the
standard model, one could be lead to the extreme point of view that
the observational lack of a Planck size cosmological constant  is
phenomenological evidence for such a UV completion.

\begin{acknowledgments}

RVB acknowledges support from DOE grant at ASU and from Arizona State
Foundation. The work of TWK was supported by DOE grant number
DE-FG05-85ER40226.

\end{acknowledgments}


\begin{thebibliography}{100}

\bibitem{Weinberg:1988cp}
  S.~Weinberg,
  Rev.\ Mod.\ Phys.\  {\bf 61}, 1 (1989).
  
\bibitem{Carroll:2000fy}
  S.~M.~Carroll,
  Living Rev.\ Rel.\  {\bf 4}, 1 (2001)
  [arXiv:astro-ph/0004075], \\
  http://relativity.livingreviews.org/Articles/lrr-2001-1/

\bibitem{Padmanabhan:2002ji}
  T.~Padmanabhan,
  Phys.\ Rept.\  {\bf 380}, 235 (2003)
  [arXiv:hep-th/0212290].
  
\bibitem{Nobbenhuis:2004wn}
 S.~Nobbenhuis,
 Found.\ Phys.\  {\bf 36}, 613 (2006)
 [arXiv:gr-qc/0411093].
 
\bibitem{Nobbenhuis:2006yf}
 S.~Nobbenhuis,
 Ph.D. Thesis, Utrecht University,
 arXiv:gr-qc/0609011.

\bibitem{Rubakov:1983bz}
  V.~A.~Rubakov and M.~E.~Shaposhnikov,
  Phys.\ Lett.\  B {\bf 125}, 139 (1983).

\bibitem{Weinberg:1987dv}
  S.~Weinberg,
  Phys.\ Rev.\ Lett.\  {\bf 59}, 2607 (1987).

\bibitem{Weinberg:2000qm}
  S.~Weinberg,
  Phys.\ Rev.\  D {\bf 61}, 103505 (2000)
  [arXiv:astro-ph/0002387].

\bibitem{Tsamis:1992sx}
  N.~C.~Tsamis and R.~P.~Woodard,
  Phys.\ Lett.\  B {\bf 301}, 351 (1993).

\bibitem{Coleman:1988tj}
  S.~R.~Coleman,
  Nucl.\ Phys.\  B {\bf 310}, 643 (1988).

\bibitem{Klebanov:1988eh}
  I.~R.~Klebanov, L.~Susskind and T.~Banks,
  Nucl.\ Phys.\  B {\bf 317}, 665 (1989).
  
\bibitem{Banks:1988je}
  T.~Banks,
  Nucl.\ Phys.\  B {\bf 309}, 493 (1988).
  
\bibitem{Carlini:1992wh}
  A.~Carlini and M.~Martellini,
  Phys.\ Lett.\  B {\bf 276}, 36 (1992)
  [arXiv:hep-th/9201042].

\bibitem{Ng:1990rw}
  Y.~J.~Ng and H.~van Dam,
  Phys.\ Rev.\ Lett.\  {\bf 65}, 1972 (1990).

\bibitem{Moffat:1996yu}
  J.~W.~Moffat,
  arXiv:astro-ph/9606071.
  
\bibitem{Dolgov:1996zg}
  A.~D.~Dolgov,
  Phys.\ Rev.\  D {\bf 55}, 5881 (1997)
  [arXiv:astro-ph/9608175].
  
\bibitem{Rubakov:1999bw}
  V.~A.~Rubakov and P.~G.~Tinyakov,
  Phys.\ Rev.\  D {\bf 61}, 087503 (2000)
  [arXiv:hep-ph/9906239].
  
\bibitem{Emelyanov:2011wn} 
  V.~Emelyanov and F.~R.~Klinkhamer,
  Phys.\ Rev.\ D {\bf 85}, 063522 (2012)
  [arXiv:1107.0961 [hep-th]].

\bibitem{Mannheim:1999nc}
  P.~D.~Mannheim,
  Astrophys.\ J.\  {\bf 561}, 1 (2001)
  [arXiv:astro-ph/9910093].
  
\bibitem{ArmendarizPicon:2000dh}
  C.~Armendariz-Picon, V.~F.~Mukhanov and P.~J.~Steinhardt,
  Phys.\ Rev.\ Lett.\  {\bf 85}, 4438 (2000)
  [arXiv:astro-ph/0004134].

\bibitem{ArmendarizPicon:2000ah}
  C.~Armendariz-Picon, V.~F.~Mukhanov and P.~J.~Steinhardt,
  Phys.\ Rev.\  D {\bf 63}, 103510 (2001)
  [arXiv:astro-ph/0006373].
  
\bibitem{Barger:2000jg}
  V.~D.~Barger and D.~Marfatia,
  Phys.\ Lett.\  B {\bf 498}, 67 (2001)
  [arXiv:astro-ph/0009256].

  
\bibitem{Chiba:2002mw}
  T.~Chiba,
  Phys.\ Rev.\  D {\bf 66}, 063514 (2002)
  [arXiv:astro-ph/0206298].

\bibitem{Chimento:2003zf}
  L.~P.~Chimento and A.~Feinstein,
  Mod.\ Phys.\ Lett.\  A {\bf 19}, 761 (2004)
  [arXiv:astro-ph/0305007].

\bibitem{Malquarti:2003nn}
  M.~Malquarti, E.~J.~Copeland, A.~R.~Liddle and M.~Trodden,
  Phys.\ Rev.\  D {\bf 67}, 123503 (2003)
  [arXiv:astro-ph/0302279].

\bibitem{Scherrer:2004au}
  R.~J.~Scherrer,
  Phys.\ Rev.\ Lett.\  {\bf 93}, 011301 (2004)
  [arXiv:astro-ph/0402316].

\bibitem{Sen:2005ra}
  A.~A.~Sen,
  JCAP {\bf 0603}, 010 (2006)
  [arXiv:astro-ph/0512406].

\bibitem{Das:2006cm}
  R.~Das, T.~W.~Kephart and R.~J.~Scherrer,
  Phys.\ Rev.\  D {\bf 74}, 103515 (2006)
  [arXiv:gr-qc/0609014].

\bibitem{'tHooft:2006rs}
 G.~'t Hooft and S.~Nobbenhuis,
 Class.\ Quant.\ Grav.\  {\bf 23}, 3819 (2006)
 [arXiv:gr-qc/0602076].

\bibitem{Bousso:2000xa}
  R.~Bousso and J.~Polchinski,
  JHEP {\bf 0006}, 006 (2000)
  [arXiv:hep-th/0004134].
   
\bibitem{Tye:2000fw}
  S.~H.~H.~Tye and I.~Wasserman,
  Phys.\ Rev.\ Lett.\  {\bf 86}, 1682 (2001)
  [arXiv:hep-th/0006068].
  
\bibitem{Kehagias:2004fb}
  A.~Kehagias,
  Phys.\ Lett.\  B {\bf 600}, 133 (2004)
  [arXiv:hep-th/0406025].
  
\bibitem{Diakonos:2007au}
  F.~K.~Diakonos and E.~N.~Saridakis,
  JCAP {\bf 0902}, 030 (2009)
  [arXiv:0708.3143 [hep-th]].
  
\bibitem{Douglas:2003um}
  M.~R.~Douglas,
  JHEP {\bf 0305}, 046 (2003)
  [arXiv:hep-th/0303194].

\bibitem{Kehagias:2000dg}
  A.~Kehagias and K.~Tamvakis,
  Mod.\ Phys.\ Lett.\  A {\bf 17}, 1767 (2002)
  [arXiv:hep-th/0011006].

\bibitem{Kim:2001ez}
  J.~E.~Kim, B.~Kyae and H.~M.~Lee,
  Nucl.\ Phys.\  B {\bf 613}, 306 (2001)
  [arXiv:hep-th/0101027].

\bibitem{Cline:2002fc}
  J.~M.~Cline and H.~Firouzjahi,
  AIP Conf.\ Proc.\  {\bf 646}, 197 (2002)
  [arXiv:hep-th/0207155].

\bibitem{Kim:2002fd}
  J.~E.~Kim,
  JHEP {\bf 0301}, 042 (2003)
  [arXiv:hep-th/0210117].
  
\bibitem{Kim:2003qy}
  J.~E.~Kim, B.~Kyae and Q.~Shafi,
  Phys.\ Rev.\  D {\bf 70}, 064039 (2004)
  [arXiv:hep-th/0305239].
  
\bibitem{Prokopec:2006yh}
  T.~Prokopec,
  arXiv:gr-qc/0603088.

\bibitem{Moffat:2002ju}
  J.~W.~Moffat,
  arXiv:hep-th/0210140.
 
\bibitem{Carter:2005sd}
  B.~M.~N.~Carter and I.~P.~Neupane,
  Phys.\ Lett.\  B {\bf 638}, 94 (2006)
  [arXiv:hep-th/0510109].

\bibitem{Afshordi:2008xu}
  N.~Afshordi,
  arXiv:0807.2639 [astro-ph].

\bibitem{Stefancic:2008zz}
  H.~Stefancic,
  Phys.\ Lett.\  B {\bf 670}, 246 (2009)
  [arXiv:0807.3692 [gr-qc]].



 \end{thebibliography}
\end{document}